\documentclass[aps,prd,preprint,tightenlines,nofootinbib,showpacs,showkeys]{revtex4}

\usepackage{graphicx}
\usepackage{dcolumn}
\usepackage{bm}

\begin{document}

\preprint{CLNS 05/1912}

\title{\boldmath Simultaneous Least Squares Treatment of Statistical and
Systematic Uncertainties}

\author{Werner M.~Sun}
\email[email: ]{wsun@mail.lepp.cornell.edu}
\affiliation{Cornell University, Ithaca, New York 14853, USA}

\date{December 19, 2005}

\begin{abstract}
We present a least squares method for estimating parameters from
measurements of event yields in the presence of background and crossfeed.
We adopt a unified approach to incorporating the statistical and systematic
uncertainties on the experimental measurements input to the fit.  We
demonstrate this method with a fit for absolute hadronic $D$ meson
branching fractions, measured in $e^+e^-\to\psi(3770)\to D\bar D$ transitions.
\end{abstract}

\maketitle

\section{Introduction}\label{sec:intro}

Least squares fitting is a well-known and powerful method for combining
information from a set of related experimental measurements to estimate the
underlying theoretical parameters (see, for instance, Reference~\cite{pdg}).
We discuss a specific
implementation of this method for use in high-energy physics experiments,
where the free parameters, denoted by the vector $\mathbf{m}$, are extracted
from event yields for signal processes.  Typically, these yields are subject
to corrections for background, crossfeed, and efficiency.  Because the sizes
of these corrections depend on the values of the free parameters, we make all
yield adjustments directly in the fit.  Often, the uncertainties on these
corrections are ignored during the fit and are propagated to the free
parameters afterwards.
However, if these uncertainties modify the relative weights of the
measurements, then the above two-step procedure would bias both the fitted
central values and the estimated uncertainties.
Therefore, we build the $\chi^2$ variable from a
full description of the uncertainties, statistical and systematic, as well
as their correlations, on both the yields and their corrections.
Thus, the input measurements --- event yields, signal efficiencies,
parameters quantifying the background processes, and background efficiencies
--- and their uncertainties are all treated in a uniform fashion.
In the $\chi^2$ minimization, we account for the $\mathbf{m}$
dependence of the yield corrections.

\section{Formalism}\label{sec:formalism}

Below, we denote matrices by upper case bold letters and one-dimensional
vectors by lower case bold letters.
Let $\mathbf{n}$ represent a set of $N$ event yield measurements,
each for a different signal process.
Each measurement may receive crossfeed contributions from other
signal processes as well as backgrounds from non-signal sources.
The background processes are described by $\mathbf{b}$, a vector of $B$
estimated
production yields, which can be functions of experimentally measured
quantities, such as branching fractions, cross sections, and luminosities.
In principle, the free parameters $\mathbf{m}$ can also appear in $\mathbf{b}$,
although no additional degrees of freedom are introduced by $\mathbf{b}$.
The rates at which these background processes contaminate the signal yields
are given by the $N\times B$ background efficiency matrix, $\mathbf{F}$.
Thus, the vector $\mathbf{s}\equiv \mathbf{n} - \mathbf{Fb}$ represents
the background-subtracted yields.

We use an $N\times N$ signal efficiency matrix, $\mathbf{E}$, to
describe simultaneously detection efficiencies (diagonal elements) and
crossfeed probabilities (off-diagonal elements).  The elements $E_{ij}$ are
defined to be the probabilities that an event of signal process $j$
is reconstructed and counted in yield $i$.  The corrected yields, denoted by
$\mathbf{c}$, are obtained by acting on $\mathbf{s}$ with the inverse of
$\mathbf{E}$:
\begin{equation}\label{eq:correctedYields}
\mathbf{c} = \mathbf{E}^{-1} \mathbf{s} =
\mathbf{E}^{-1}( \mathbf{n} - \mathbf{Fb} ).
\end{equation}
Thus, $\mathbf{c}$ encapsulates all the experimental measurements.
The variance matrix of $\mathbf{c}$, denoted by $\mathbf{V_c}$, receives
contributions,
both statistical and systematic, from each element of $\mathbf{n}$,
$\mathbf{b}$, $\mathbf{E}$, and $\mathbf{F}$.  

In the least squares fit, we define
$\chi^2 \equiv \left(\mathbf{c}-\mathbf{\widetilde c}\right)^T \mathbf{V}_{\mathbf{c}}^{-1} \left(\mathbf{c}-\mathbf{\widetilde c}\right)$,
where $\mathbf{\widetilde c}$ is the vector of predicted yields, which are
also functions of $\mathbf{m}$.
Because both $\mathbf{\widetilde c}$ and $\mathbf{c}$ (through $\mathbf{b}$)
depend on $\mathbf{m}$, minimizing this $\chi^2$ amounts to a nonlinear
version of the total least squares method~\cite{tls}.  We solve this problem
by extending the conventional least squares fit to include contributions from
both $\mathbf{\widetilde c}$ and $\mathbf{c}$ in
$\partial\chi^2/\partial\mathbf{m}$.  Given a set of seed values,
$\mathbf{m}_0$, the optimized estimate,
$\mathbf{\widehat m}$, and its variance matrix, $\mathbf{V_m}$, are
\begin{eqnarray}
\label{eq:fittedParameters}
\mathbf{\widehat m} &=& \mathbf{m}_0 +
	\left(\mathbf{D}\mathbf{V}_{\mathbf{c}}^{-1}\mathbf{D}^T\right)^{-1}
	\mathbf{D}\mathbf{V}_{\mathbf{c}}^{-1}
	\left[\mathbf{c}(\mathbf{m}_0)-\mathbf{\widetilde c}(\mathbf{m}_0)\right]\\
\label{eq:fittedError}
\mathbf{V_m} &=& \frac{1}{2}\frac{\partial^2\chi^2}{\partial\mathbf{m}\,
	\partial\mathbf{m}^T} =
	\left(\mathbf{D}\mathbf{V}_{\mathbf{c}}^{-1}\mathbf{D}^T\right)^{-1},
\end{eqnarray}
where the $M\times N$ derivative matrix $\mathbf{D}$ is defined to be
\begin{equation}
\mathbf{D}\equiv
\frac{\partial\mathbf{\widetilde c}}{\partial\mathbf{m}} -
\frac{\partial\mathbf{c}}{\partial\mathbf{m}} =
\frac{\partial\mathbf{\widetilde c}}{\partial\mathbf{m}} +
\frac{\partial\mathbf{b}}{\partial\mathbf{m}}
\mathbf{F}^T\left(\mathbf{E}^{-1}\right)^T.
\end{equation}
In general, $\mathbf{\widetilde c}$ and $\mathbf{c}$ are nonlinear
functions of $\mathbf{m}$, so the linearized solution
$\mathbf{\widehat m}$ is approximate, and the above procedure is
iterated until the $\chi^2$ converges. Between iterations, all the
fit inputs that depend on $\mathbf{m}$ are reevaluated with the updated
values of $\mathbf{\widehat m}$.

Nonlinearities also occur when $\mathbf{V_c}$ contains multiplicative
or Poisson uncertainties that depend on the measurement values.
With the least squares method, these nonlinearities result in biased
estimators unless these variable uncertainties are
evaluated using the predicted yields $\mathbf{\widetilde c}$ instead of the
measured $\mathbf{c}$.  Therefore, all three ingredients in the $\chi^2$ ---
$\mathbf{c}$, $\mathbf{\widetilde c}$, and $\mathbf{V_c}$ --- are functions of
$\mathbf{m}$.  However, we do not include the derivatives
$\partial\mathbf{V_c}/\partial\mathbf{m}$ in $\mathbf{D}$ because
doing so would generate biases in $\mathbf{\widehat m}$.

For a simple demonstration of the aforementioned biases, we
consider two measured yields, $c_1$ and $c_2$, which are both estimators of
a true yield $\bar c$.  We assume that the uncertainties on $c_1$ and $c_2$
are uncorrelated, multiplicative, and of the same fractional size, $\lambda$.
We construct an improved estimator, $\widehat c$, by minimizing
$\chi^2 = (c_1-c)^2/\sigma_{c_1}^2 + (c_2-c)^2/\sigma_{c_2}^2$ with respect to
$c$.  If, following the prescription given above, we neglect the
$\partial\sigma_{c_i}^2/\partial c$ terms in
$\partial\chi^2/\partial c$ and assign (iteratively) the uncertainties
$\sigma_{c_1}=\sigma_{c_2}=\lambda\widehat c$,
then $c_1$ and $c_2$ are equally weighted, and $\widehat c$ is an unbiased
estimate of $\bar c$:
\begin{eqnarray}
\widehat c_{\rm unbiased} &=& \frac{c_1+c_2}{2} \\
\chi^2_{\rm unbiased} &=&
	\frac{2}{\lambda^2}\left(\frac{c_1-c_2}{c_1+c_2}\right)^2 .
\end{eqnarray}
On the other hand, including the $\partial\sigma_{c_i}^2/\partial c$ terms in
$\partial\chi^2/\partial c$ results in an upward bias:
\begin{eqnarray}
\widehat c_{\rm biased1} &=& \frac{c_1^2+c_2^2}{c_1+c_2} =
	\widehat c_{\rm unbiased}\left(1+\frac{\lambda^2\chi^2_{\rm unbiased}}{2}\right) \\
\chi^2_{\rm biased1} &=& \frac{(c_1-c_2)^2}{\lambda^2(c_1^2+c_2^2)} .
\end{eqnarray}
Finally, if we assign uncertainties based on the measured yields, not the
predicted yields, such that $\sigma_{c_1}=\lambda c_1$,
$\sigma_{c_2}=\lambda c_2$, and $\partial\sigma_{c_i}^2/\partial c=0$,
then the resulting estimate is biased low:
\begin{eqnarray}
\widehat c_{\rm biased2} &=& \frac{c_1 c_2 (c_1+c_2)}{c_1^2+c_2^2} =
	\widehat c_{\rm unbiased}(1-\lambda^2\chi^2_{\rm biased1}) \\
\chi^2_{\rm biased2} &=& \chi^2_{\rm biased1}.
\end{eqnarray}
Thus, even though $\chi^2_{\rm biased1}$ and $\chi^2_{\rm biased2}$ are
smaller than $\chi^2_{\rm unbiased}$, the corresponding estimators possess
undesired properties.

\section{\boldmath Input Variance Matrix}
\label{sec:inputVarianceMatrix}

The uncertainties on the $N$ elements of $\mathbf{n}$ and the $B$ elements
of $\mathbf{b}$ are characterized by the $N\times N$ matrix
$\mathbf{V_n}$ and the $B\times B$ matrix $\mathbf{V_b}$, respectively.
Usually, the elements of $\mathbf{E}$ and $\mathbf{F}$ share many common
correlated systematic uncertainties, so we construct a joint
variance matrix from the submatrices $\mathbf{V_E}$, $\mathbf{V_F}$, and
$\mathbf{C_{EF}}$, where
$\mathbf{V_E}$ ($N^2\times N^2$) and $\mathbf{V_F}$ ($NB\times NB$) are
the variance matrices for the elements of $\mathbf{E}$ and $\mathbf{F}$,
respectively, and $\mathbf{C_{EF}}$ ($N^2\times NB$)
contains the correlations between $\mathbf{E}$ and $\mathbf{F}$.
Below, we label each element of $\mathbf{E}$
or $\mathbf{F}$ by two indices ($E_{ij}$ or $F_{ij}$), and the two dimensions
of $\mathbf{E}$ or $\mathbf{F}$ are mapped onto one dimension of
$\mathbf{V_E}$ or $\mathbf{V_F}$.

We form $\mathbf{V_c}$ by propagating the statistical and systematic
uncertainties on $\mathbf{n}$, $\mathbf{b}$, $\mathbf{E}$, and $\mathbf{F}$
to $\mathbf{c}$ via
\begin{equation}
\label{eq:errorPropagation1}
\mathbf{V_c} =
\frac{\partial\mathbf{c}}{\partial\mathbf{n}}^T \mathbf{V_n}
\frac{\partial\mathbf{c}}{\partial\mathbf{n}} +
\frac{\partial\mathbf{c}}{\partial\mathbf{b}}^T \mathbf{V_b}
\frac{\partial\mathbf{c}}{\partial\mathbf{b}} +
\left(\begin{array}{cc}
(\partial\mathbf{c}/\partial\mathbf{E})^T &
(\partial\mathbf{c}/\partial\mathbf{F})^T
\end{array}\right)
\left(\begin{array}{cc}
\mathbf{V_E} & \mathbf{C_{EF}} \\
\mathbf{C}_{\mathbf{EF}}^T & \mathbf{V_F}
\end{array}\right)
\left(\begin{array}{c}
\partial\mathbf{c}/\partial\mathbf{E} \\
\partial\mathbf{c}/\partial\mathbf{F}
\end{array}\right).
\end{equation}
Where appropriate, we substitute $\mathbf{\widetilde c}$ for $\mathbf{c}$, as
discussed in Section~\ref{sec:formalism}.
The first term of Equation~\ref{eq:errorPropagation1} is simply
$\mathbf{E}^{-1}\mathbf{V_n} (\mathbf{E}^{-1})^T$, and the
second term is
$\mathbf{E}^{-1}\mathbf{F}\mathbf{V_b}\mathbf{F}^T (\mathbf{E}^{-1})^T$.
For the third term, we evaluate the partial derivatives and find
\begin{eqnarray}
\frac{\partial\mathbf{c}}{\partial\mathbf{E}} &=&
	\mathbf{s}^T
	\left(\frac{\partial\mathbf{E}^{-1}}{\partial\mathbf{E}}\right)^T =
	-\mathbf{s}^T \left(\mathbf{E}^{-1}\right)^T
	\left(\frac{\partial\mathbf{E}}{\partial\mathbf{E}}\right)^T
	\left(\mathbf{E}^{-1}\right)^T =
	-\mathbf{A}\left(\mathbf{E}^{-1}\right)^T \\
\frac{\partial\mathbf{c}}{\partial\mathbf{F}} &=&
	-\mathbf{B}\left(\mathbf{E}^{-1}\right)^T,
\end{eqnarray}
where
$\mathbf{A}\equiv\mathbf{c}^T (\partial\mathbf{E}/\partial\mathbf{E})^T$ and
$\mathbf{B}\equiv\mathbf{b}^T (\partial\mathbf{F}/\partial\mathbf{F})^T$, with
elements given in terms of the Kronecker delta ($\delta_{ij}$):
$\partial E_{kl}/\partial E_{ij}=\partial F_{kl}/\partial F_{ij}=\delta_{ik}\delta_{jl}$.
The matrices $\mathbf{A}$ and $\mathbf{B}$ have rows labeled by two indices,
which refer to the elements of $\mathbf{E}$ and $\mathbf{F}$, respectively,
and columns labeled by one index, which refers to the elements of $\mathbf{c}$.
In other words, the $ij$-th row of $\mathbf{A}$ is given by 
$\mathbf{c}^T (\partial\mathbf{E}/\partial E_{ij})^T$, where
$(\partial\mathbf{E}/\partial E_{ij})_{kl} = \partial E_{kl}/\partial E_{ij}$.
Therefore, the elements of $\mathbf{A}$ and $\mathbf{B}$ are
$A_{ij,k} = \delta_{ik} \widetilde c_j$ and
$B_{ij,k} = \delta_{ik} b_j$.
For $N=B=2$, these matrices are
\begin{equation}
\label{eq:Adefinition}
\mathbf{A} =
\left(\begin{array}{cc}
\widetilde c_1 & 0 \\
\widetilde c_2 & 0 \\
0 & \widetilde c_1 \\
0 & \widetilde c_2 \\
\end{array}\right)
\hspace{1cm}{\rm and}\hspace{1cm}
\mathbf{B} =
\left(\begin{array}{cc}
b_1 & 0 \\
b_2 & 0 \\
0 & b_1 \\
0 & b_2 \\
\end{array}\right).
\end{equation}
This treatment of error propagation in matrix inversion agrees with
that derived in Reference~\cite{Lefebvre:1999yu}.
The above relations allow us to reexpress $\mathbf{V_c}$ as
\begin{equation}
\label{eq:errorPropagation2}
\mathbf{V_c} = \mathbf{E}^{-1}\mathbf{V_{\Delta n}}
\left(\mathbf{E}^{-1}\right)^T,
\end{equation}
where $\mathbf{V_{\Delta n}}\equiv \mathbf{V_n} + \mathbf{F}\mathbf{V_b}\mathbf{F}^T + \mathbf{A}^T \mathbf{V_E} \mathbf{A} + \mathbf{B}^T \mathbf{V_F} \mathbf{B} + \mathbf{A}^T \mathbf{C_{EF}} \mathbf{B} + \mathbf{B}^T \mathbf{C}_{\mathbf{EF}}^T \mathbf{A}$.
As a result, we have
$\chi^2 = \mathbf{\Delta n}^T \mathbf{V}_{\mathbf{\Delta n}}^{-1}\mathbf{\Delta n}$,
where $\mathbf{\Delta n}\equiv\mathbf{n}-\mathbf{E\widetilde c}-\mathbf{Fb}$.
Thus, the $\chi^2$ minimization can be formulated equivalently in terms of
$\mathbf{n}$ instead of $\mathbf{c}$:
$\mathbf{V_m} = \left(\mathbf{D'}\mathbf{V}_{\mathbf{\Delta n}}^{-1}\mathbf{D'}^T\right)^{-1}$ and
$\mathbf{\widehat m} = \mathbf{m}_0 + \mathbf{V_m} \mathbf{D'}\mathbf{V}_{\mathbf{\Delta n}}^{-1} \mathbf{\Delta n}$,
where $\mathbf{D'}\equiv \mathbf{D}\mathbf{E}^T = (\partial\mathbf{\widetilde c}/\partial\mathbf{m})\mathbf{E}^T +(\partial\mathbf{b}/\partial\mathbf{m})\mathbf{F}^T$.

Systematic uncertainties on the efficiencies are often
multiplicative and belong to one of three categories: those that depend
only on the reconstructed mode (row-wise),
those that depend only on the generated mode (column-wise),
and those that are uncorrelated among elements of
$\mathbf{E}$ and $\mathbf{F}$.
For row-wise efficiency uncertainties, all the elements in any given
row of $\mathbf{E}$ and $\mathbf{F}$ have the same fractional uncertainty,
which we denote by
$\lambda_i \equiv \sigma_{E_{ij}}/E_{ij} = \sigma_{F_{ij}}/F_{ij}$.
The correlation coefficients between elements of different rows are
$\lambda_{ij}^2 / (\lambda_i\lambda_j)$,
where $\lambda_{ij}$ characterizes the uncertainties common to
$c_i$ and $c_j$.  For instance, if $\lambda_{\rm track}$ is the fractional
uncertainty associated with the charged particle tracking efficiency, then
$\lambda_i = t_i\lambda_{\rm track}$ and
$\lambda_{ij}^2 = t_i t_j \lambda_{\rm track}^2$, where
$t_i$ and $t_j$ are the track multiplicities in modes $i$ and $j$,
respectively.  Note that $\lambda_{ii} = \lambda_i$.  Similarly, for
column-wise uncertainties, we define the fractional uncertainties
$\mu_j\equiv \sigma_{E_{ij}}/E_{ij} = \sigma_{F_{ij}}/F_{ij}$ and correlation
coefficients $\mu_{ij}^2/(\mu_i\mu_j)$.  We denote the uncorrelated
fractional uncertainty on any element of $\mathbf{E}$ or $\mathbf{F}$
by $\nu_{ik, jl}$.  Table~\ref{tab:vefElements} gives expressions
for the elements of $\mathbf{V_E}$, $\mathbf{V_F}$, and $\mathbf{C_{EF}}$,
as well as their contributions to $\mathbf{V_c}$ for row-wise, column-wise,
and uncorrelated uncertainties.

\begin{table}[ht]
\begin{center}
\caption{Expressions for the elements of $\mathbf{V_E}$, $\mathbf{V_F}$,
and $\mathbf{C_{EF}}$, as well as their contributions to $\mathbf{V_c}$.
Repeated external indices are not summed over.}
\label{tab:vefElements}
\begin{tabular}{cccc}
\hline\hline
Quantity & Row-wise & Column-wise & Uncorrelated \\
\hline
$(\mathbf{V_E})_{ik, jl}$ &
	$\lambda_{ij}^2 E_{ik}E_{jl}$ &
	$\mu_{kl}^2 E_{ik}E_{jl}$ &
	$\nu_{ik,jl}^2 E_{ik}E_{jl}\delta_{ij}\delta_{kl}$ \\
$(\mathbf{V_F})_{ik, jl}$ &
	$\lambda_{ij}^2 F_{ik}F_{jl}$ &
	$\mu_{kl}^2 F_{ik}F_{jl}$ &
	$\nu_{ik,jl}^2 F_{ik}F_{jl}\delta_{ij}\delta_{kl}$ \\
$(\mathbf{C_{EF}})_{ik, jl}$ &
	$\lambda_{ij}^2 E_{ik}F_{jl}$ &
	$\mu_{kl}^2 E_{ik}F_{jl}$ &
	0 \\
\hline
$(\mathbf{A}^T \mathbf{V_E} \mathbf{A})_{ij}$ &
	$\lambda_{ij}^2 \widetilde s_i \widetilde s_j$ &
	$\mu_{kl}^2 E_{ik} \widetilde c_k E_{jl} \widetilde c_l$ &
	$\delta_{ij} \sigma^2_{E_{jk}} \widetilde c_k^2$ \\
$(\mathbf{B}^T \mathbf{V_F} \mathbf{B})_{ij}$ &
	$\lambda_{ij}^2 F_{ik}b_k F_{jl}b_l$ &
	$\mu_{kl}^2 F_{ik}b_k F_{jl}b_l$ &
	$\delta_{ij} \sigma_{F_{jk}}^2 b_k^2$ \\
$(\mathbf{A}^T \mathbf{C_{EF}} \mathbf{B})_{ij}$&
	$\lambda_{ij}^2 \widetilde s_i F_{jk}b_k$ &
	$\mu_{kl}^2 E_{ik} \widetilde c_k F_{jl}b_l$ &
	0 \\
\hline\hline
\end{tabular}
\end{center}
\end{table}

\section{\boldmath Example: Hadronic $D$ Meson Branching Fractions}

The least squares method described in the previous sections has been employed
by the CLEO-c collaboration~\cite{cleoc-dhad} to measure absolute branching
fractions for hadronic $D$ meson decays.  Using $D\bar D$ pairs produced
through the $\psi(3770)$ resonance, the branching fraction for mode $i$,
denoted by ${\cal B}_i$, is measured by comparing the number of events
where a single $D\to i$ decay is reconstructed (called single tag,
denoted by $x_i$) with the number of events where both $D$ and $\bar D$ are
reconstructed via $D\to i$ and $\bar D\to j$ (called double tag, denoted by
$y_{ij}$).  These yield measurements form the vector $\mathbf{n}$.
The free parameters $\mathbf{m}$ are the ${\cal B}_i$
and the numbers of $D^0\bar D^0$ and $D^+D^-$ pairs produced, denoted by
${\cal N}^{00}$ and ${\cal N}^{+-}$, respectively, and denoted generically
by ${\cal N}$.
Yields for charge conjugate modes are measured separately, so the predicted
corrected yields $\mathbf{\widetilde c}$ are ${\cal N}{\cal B}_i$ for single
tags and ${\cal N}{\cal B}_i{\cal B}_j$ for double tags.
Thus, ${\cal B}_i$ and ${\cal N}$ can be
extracted from various products and ratios of $x_i$, $x_j$, and $y_{ij}$:
${\cal B}_i \sim y_{ij}/x_j$, ${\cal N}\sim x_i x_j/y_{ij}$, up to
corrections for efficiency, crossfeed, and background.

The matrix $\mathbf{V_n}$ describes the statistical uncertainties and
correlations among the $x_i$ and $y_{ij}$.  The $y_{ij}$ are uncorrelated,
but because any given event can contain both single tag and double tag
candidates, the $x_i$ are correlated among themselves as well as with the
$y_{ij}$.  If the selection criteria for single and double tags are the same,
then the events (signal and background) used to estimate $y_{ij}$ are a proper
subset of those for $x_i$ and $x_j$.  Thus, any single tag yield is a sum of
exclusive single tags ($x_i^{\rm excl}$) and double tags:
$x_{\{i,j\}} = x_{\{i,j\}}^{\rm excl} + y_{ij}$.
Propagating the uncertainties on the independent variables,
$x_i^{\rm excl}$, $x_j^{\rm excl}$, and $y_{ij}$, gives the following elements
for $\mathbf{V_n}$:
\begin{eqnarray}
\label{eq:singleVar}
\langle \Delta x_i \Delta x_j \rangle &=&
	\delta_{ij} \sigma_{x_i} \sigma_{x_j} +
	(1-\delta_{ij}) \sigma^2_{y_{ij}} \\
\label{eq:doubleVar}
\langle \Delta y_{ij} \Delta y_{kl} \rangle
	&=& \delta_{ik}\delta_{jl}\sigma_{y_{ij}}\sigma_{y_{kl}} \\
\label{eq:singleDoubleCovar}
\langle \Delta x_i \Delta y_{jk} \rangle &=&
	(\delta_{ij} + \delta_{ik}) \sigma^2_{y_{jk}},
\end{eqnarray}
where $\Delta x_i\equiv x_i - \langle x_i\rangle$,
$\Delta y_{ij} \equiv y_{ij} - \langle y_{ij}\rangle$, and
$\sigma_{x_{\{i,j\}}}^2=\sigma_{x_{\{i,j\}}^{\rm excl}}^2 + \sigma_{y_{ij}}^2$.
Thus, for any two single tag yields and the corresponding double tag
yield, the three off-diagonal elements of $\mathbf{V_n}$ are all
given by the uncertainty on the number of overlapping events.
In addition to these statistical uncertainties,
$\mathbf{V_n}$ can also receive contributions from additive systematic
uncertainties.

Some of the sources of background we consider are non-signal $D$ decays,
$e^+e^-\to q\bar q$ events, and $e^+e^-\to\tau^+\tau^-$ events.
If there are two non-signal $D$ backgrounds with branching fractions
${\cal C}_1$ and ${\cal C}_2$, then the vector $\mathbf{b}$ is given by
\begin{equation}\label{eq:backrounds}
\mathbf{b} = \left(\begin{array}{c}
{\cal N}{\cal C}_1 \\
{\cal N}{\cal C}_2 \\
{\cal L} X_{q\bar q} \\
{\cal L} X_{\tau^+\tau^-}
\end{array}\right),
\end{equation}
where $X_{q\bar q}$ and $X_{\tau^+\tau^-}$ are the cross sections
for $q\bar q$ and $\tau^+\tau^-$ production, respectively, and ${\cal L}$ is
the integrated luminosity of the data sample.
Because of the non-signal $D$ decays, the free parameter ${\cal N}$ appears in
$\mathbf{b}$ but does not contribute any additional terms to the variance
matrix $\mathbf{V_b}$, which takes the following block diagonal form:
\begin{equation}
\mathbf{V_b} = \left(\begin{array}{cccc}
{\cal N}^2\sigma_{{\cal C}_1}^2 & 0 & 0 & 0 \\
0 & {\cal N}^2\sigma_{{\cal C}_2}^2 & 0 & 0 \\
0 & 0 & {\cal L}^2\sigma_{X_{q\bar q}}^2 + X_{q\bar q}^2\sigma_{\cal L}^2 &
	X_{q\bar q} X_{\tau^+\tau^-} \sigma_{\cal L}^2 \\
0 & 0 & X_{q\bar q} X_{\tau^+\tau^-} \sigma_{\cal L}^2 &
	{\cal L}^2\sigma_{X_{\tau^+\tau^-}}^2 + X_{\tau^+\tau^-}^2
	\sigma_{\cal L}^2
\end{array}\right).
\end{equation}
 Also, the matrix
$\partial\mathbf{b}/\partial\mathbf{m}$ is nontrivial and is incorporated into
the $\chi^2$ minimization.

In the joint variance matrix for $\mathbf{E}$ and $\mathbf{F}$, uncertainties
of all three types discussed in Section~\ref{sec:inputVarianceMatrix} are
present. Row-wise effects arise from systematic uncertainties on simulated
reconstruction efficiencies for charged tracks, $\pi^0\to\gamma\gamma$
decays, $K^0_S\to\pi^+\pi^-$ decays, and
particle identification (PID) for charged pions and kaons.
Column-wise uncertainties reflect the poorly known resonant substructure in
multi-body final states.  Uncorrelated contributions come from statistical
uncertainties due to the finite Monte Carlo (MC) simulated samples used to
determine $\mathbf{E}$ and $\mathbf{F}$.
Thus, for example, if mode $i$ is $D^0\to K^-\pi^+\pi^0$ and mode $j$ is
$D^+\to K^0_S\pi^+$, then the row-wise uncertainties are given by
\begin{eqnarray}
\lambda_i^2 & = & ( 2\lambda_{\rm track} )^2 + \lambda_{\pi^0}^2 +
	\lambda_{\pi^\pm {\rm PID}}^2 + \lambda_{K^\pm {\rm PID}}^2 \\
\lambda_j^2 & = & ( 3\lambda_{\rm track} )^2 +
	\lambda_{\pi^\pm {\rm PID}}^2 \\
\lambda_{ij}^2 & = & 6\lambda_{\rm track}^2 + \lambda_{\pi^\pm {\rm PID}}^2.
\end{eqnarray}
Because these row-wise and column-wise uncertainties are completely correlated
among the yields to which they pertain, they degrade the precision of
${\cal B}_i$ but not ${\cal N}$.  Furthermore, they have no effect
on the central values of $\mathbf{\widehat m}$ because
the relative weight of each yield is unaltered by these
uncertainties.  However, they can introduce large systematic correlations
among the fit parameters, even between statistically independent branching
fractions of different charge.

\subsection{Toy Monte Carlo Study}\label{sec:toyMC}

We test the method presented above using a toy MC simulation with Gaussian
smearing of the fit inputs.  We generate data for five decay modes,
$D^0\to K^-\pi^+$, $D^0\to K^-\pi^+\pi^0$, $D^0\to K^-\pi^+\pi^-\pi^+$,
$D^+\to K^-\pi^+\pi^+$, and $D^+\to K^0_S\pi^+$ (charge conjugate particles
are implied), for which there are ten single tag and thirteen double tag
yields.  The fit determines seven free parameters: ${\cal N}^{00}$,
${\cal N}^{+-}$, and five charge-averaged branching fractions.  The input
branching fractions are taken to be the world-average values given in
Reference~\cite{pdg}, and we use ${\cal N}^{00}=2.0\times 10^5$ and
${\cal N}^{+-}=1.5\times 10^5$.  The efficiencies are mode-dependent:
30\%--70\% for single tags and 10\%--50\% for double tags, with fractional
statistical uncertainties of 0.5\%--1.0\%.  The yield uncertainties are
specified to be close to the Poisson limit, and backgrounds correspond roughly
to those expected in 60 ${\rm pb}^{-1}$ of $e^+e^-$ collisions at the
$\psi(3770)$. Also, we apply correlated systematic efficiency uncertainties of
1\% for tracking, 2\% for $\pi^0$ reconstruction, 2\% for $K^0_S$
reconstruction, and 1\% for charged pion and kaon PID.

\begin{figure}
\includegraphics*[width=0.5\linewidth]{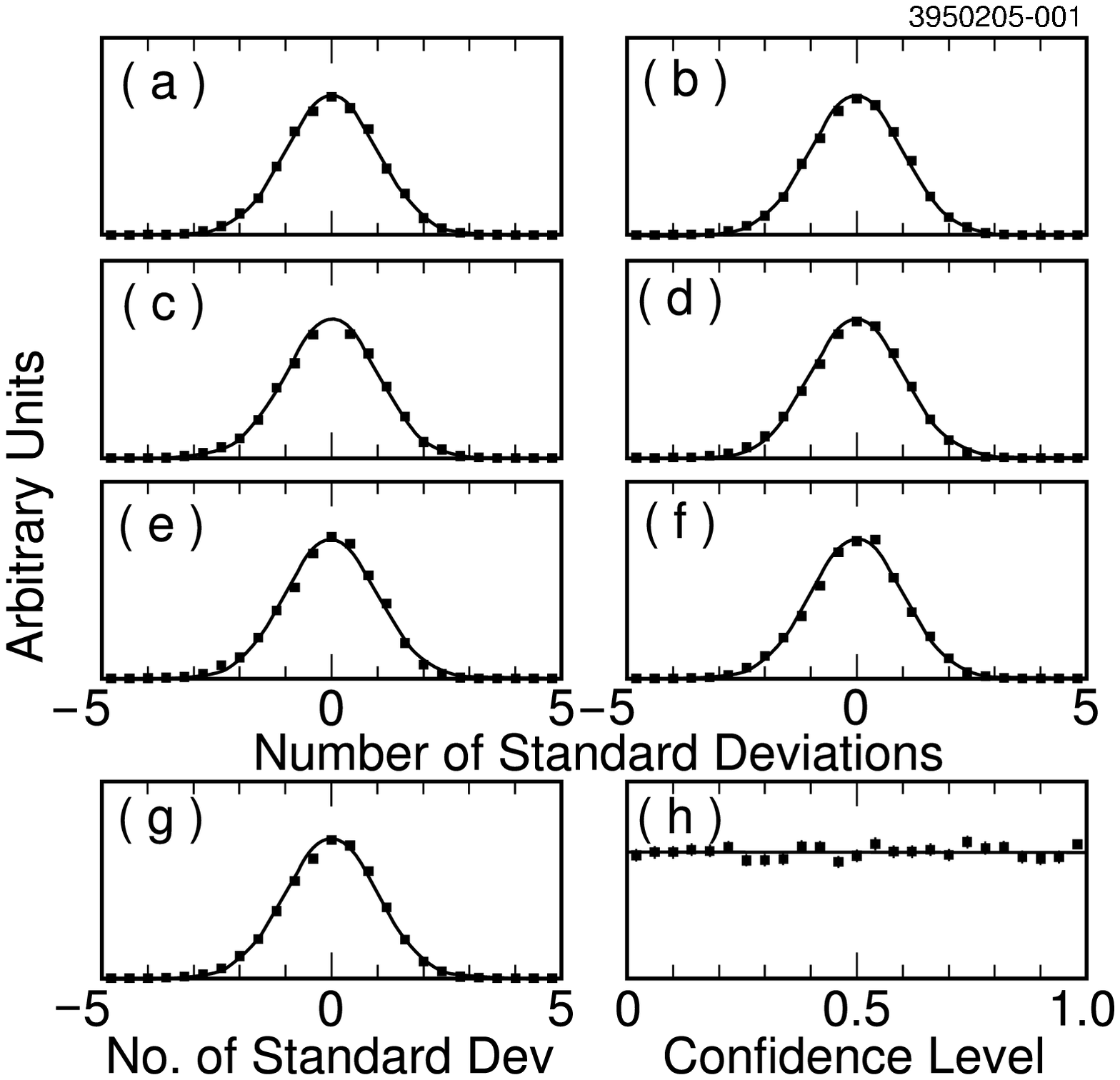}
\caption{Toy MC fit pull distributions for ${\cal N}^{00}$ (a),
${\cal B}(D^0\to K^-\pi^+)$ (b), ${\cal B}(D^0\to K^-\pi^+\pi^0)$ (c),
${\cal B}(D^0\to K^-\pi^+\pi^-\pi^+)$ (d), ${\cal N}^{+-}$ (e),
${\cal B}(D^+\to K^-\pi^+\pi^+)$ (f), and ${\cal B}(D^+\to K^0_S\pi^+)$ (g),
overlaid with Gaussian curves with zero mean and unit width.  The fit
confidence level distribution (h) is overlaid with a line with zero slope.}
\label{fig:toyMCPulls}
\end{figure}

The fit reproduces the input parameters well.
Figure~\ref{fig:toyMCPulls} shows the pull distributions for the seven
fit parameters and the fit confidence level for 10000 toy MC trials.
All the pull distributions are unbiased and have widths consistent with unity.
Also, the confidence level is flat.  Table~\ref{tab:correlations} gives the
correlation coefficients among the fit parameters.  Branching fractions tend
to be positively correlated with each other and negatively correlated with
${\cal N}^{00}$ and ${\cal N}^{+-}$.  In particular, the $D^0$ branching
fractions are correlated with those for $D^+$.
In the absence of correlated efficiency uncertainties, the $D^0$ and $D^+$
free parameters would essentially be independent.

\begin{table}[htb]
\caption{Correlation coefficients, including systematic uncertainties,
for the free parameters determined by the fit to toy MC samples.}
\label{tab:correlations}
\begin{center}
\begin{tabular}{l|ccccccc}
\hline\hline
& ~~${\cal N}^{00}$~~ & ~~$K^-\pi^+$~~ & ~~$K^-\pi^+\pi^0$~~ &
	~~$K^-\pi^+\pi^-\pi^+$~~ &
        ~~${\cal N}^{+-}$~~ & ~~$K^-\pi^+\pi^+$~~ & ~~$K^0_S\pi^+$~~ \\
\hline
${\cal N}^{00}$ & 1 & $-0.63$ & $-0.52$ & $-0.38$
        & $-0.01$ & $-0.01$ & $-0.01$ \\
$K^-\pi^+$  & & 1 & 0.79 &  0.87 & $-0.01$ & 0.40 & 0.29 \\
$K^-\pi^+\pi^0$ & & & 1 & 0.77 & $-0.01$ & 0.37 & 0.27 \\
$K^-\pi^+\pi^-\pi^+$ & & & & 1 & $-0.01$ & 0.53 & 0.39 \\ 
${\cal N}^{+-}$ & & & & & 1 & $-0.82$ & $-0.77$ \\
$K^-\pi^+\pi^+$ & & & & & & 1 & 0.87 \\
$K^0_S\pi^+$ & & & & & & & 1\\
\hline\hline
\end{tabular}
\end{center}
\end{table}

Slight asymmetries can be observed in the pull distributions, especially
in those for ${\cal N}^{00}$ and ${\cal N}^{+-}$.  These asymmetries are
caused by the nonlinear nature of the multiplicative efficiency uncertainties
and of the functions $\mathbf{\widetilde c}(\mathbf{m})$.  Because the fit
parameters are effectively estimated from ratios of the input yields, Gaussian
fluctuations in the denominators produce non-Gaussian fluctuations in the
ratios, which are most visible in ${\cal N}^{00}$ and ${\cal N}^{+-}$, where
the uncertainties in the denominators are dominant.
Similarly, multiplicative uncertainties, which affect only the
branching fractions, scale with the fitted values and, therefore, give rise to
asymmetric ${\cal B}$ pulls.  In both cases, larger fractional uncertainties
would heighten the asymmetries.

\begin{figure}
\includegraphics*[width=0.5\linewidth]{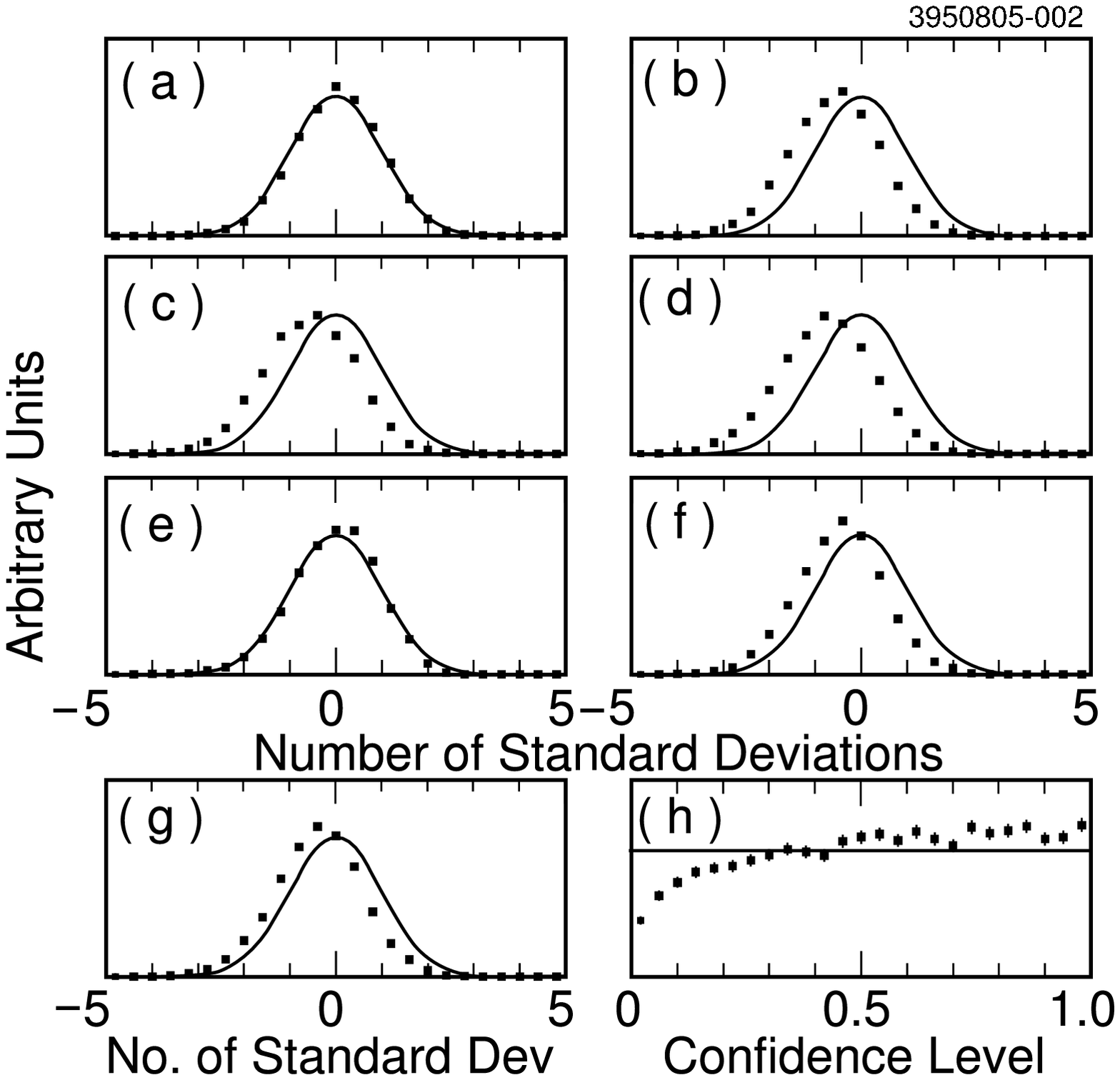}
\caption{Toy MC fit pull distributions, with $\mathbf{V_c}$ calculated using
$\mathbf{c}$ instead of $\mathbf{\widetilde c}$, for ${\cal N}^{00}$ (a),
${\cal B}(D^0\to K^-\pi^+)$ (b), ${\cal B}(D^0\to K^-\pi^+\pi^0)$ (c),
${\cal B}(D^0\to K^-\pi^+\pi^-\pi^+)$ (d), ${\cal N}^{+-}$ (e),
${\cal B}(D^+\to K^-\pi^+\pi^+)$ (f), and ${\cal B}(D^+\to K^0_S\pi^+)$ (g),
overlaid with Gaussian curves with zero mean and unit width.  The fit
confidence level distribution (h) is overlaid with a line with zero slope.}
\label{fig:toyMCPullsBiased}
\end{figure}

If we form the matrix $\mathbf{A}$ in Equation~\ref{eq:Adefinition} using the
measured yields $\mathbf{c}$ rather than the predicted yields
$\mathbf{\widetilde c}$, then the variance matrix $\mathbf{V_c}$ need not be
reevaluated after each fit iteration.  However, in this case, the pull
distributions become significantly biased, as shown in
Figure~\ref{fig:toyMCPullsBiased}.  Thus, obtaining
unbiased fit results and the correct uncertainties requires proper handling
of the efficiency variance matrices $\mathbf{V_E}$ and $\mathbf{V_F}$.

\section{Summary}

We have developed a least squares fit that simultaneously incorporates
statistical and systematic uncertainties, as well as their correlations,
on all the input
experimental measurements.  Biases from nonlinearities are reduced by
introducing fit parameter dependence in the input variance matrix.
This fitting method is used to measure absolute branching fractions of
hadronic $D$ meson decays, and toy Monte Carlo studies validate the performance
of the fitter.  By including all known sources of measurement uncertainty in
the $\chi^2$, we obtain unbiased fit parameters with correct estimated
uncertainties.

\begin{acknowledgments}
We wish to thank Roy Briere, David Cassel, Lawrence Gibbons, Wolfgang Rolke,
Anders Ryd, and Ian Shipsey for many helpful discussions.
This work was supported in part by the National Science Foundation under
Grant No. PHY-0202078.
\end{acknowledgments}

\end{document}